\documentclass[12pt]{article}
\usepackage{moreverb}
\usepackage{url}
\usepackage{grffile}
\usepackage{lineno}
\usepackage{graphics,graphicx}
\usepackage{caption}
\usepackage{bm}
\usepackage[english]{babel}
\usepackage[square,numbers]{natbib}

\usepackage{atbegshi}
\AtBeginDocument{\AtBeginShipoutNext{\AtBeginShipoutDiscard}}

\usepackage[a4paper, total={7in, 9in}]{geometry}

\usepackage{amsmath}

\DeclareMathOperator*{\argmin}{arg\,min}

\begin{document}
\title{Comparison of Azzalini and Geometric Skew Normal Distributions Under Bayesian Paradigm}
\author{Narayan Srinivasan$^\ast$}
\thanks{Project student at Indian Institute of Science Education and Research Kolkata, \\ email: \texttt{narayanseshadri98@gmail.com}}

\maketitle

\begin{abstract}
Skewed generalizations of the normal distribution have been a topic of great interest in the statistics community due to their diverse applications across several domains. One of the most popular skew normal distributions, due to its intuitive appeal, is the Azzalini's skew normal distribution. However, due to the nature of the distribution it suffers from serious inferential problems. Interestingly, the Bayesian approach has been shown to mitigate these issues. Recently, another skew normal distribution, the Geometric skew normal distribution, which is structurally different from Azzalini's skew normal distribution, has been proposed as an alternative for modelling skewed data. Despite the interest in skew normal distributions, a limited number of articles deal with comparing the performance of different skew distributions, especially in the Bayesian context. To address this gap, the article attempts to compare these two skew normal distributions in the Bayesian paradigm for modelling data with varied skewness. The posterior estimates of the parameters of the geometric skew normal distribution are obtained using a hybrid Gibbs-Metropolis algorithm and the posterior predictive fit to the data is also obtained. Similarly, for the Azzalini's skew normal distribution, the posterior predictive fit are derived using a Gibbs sampling algorithm. To compare the performance, Kolmogorov-Smirnov distance between the posterior predictive distribution and original data is utilised to measure the goodness of fit for both the models. An assortment of real and simulated data sets, with a wide range of skewness, are analyzed using both the models and the advantages as well as disadvantages of the two models are also discussed. Finally, a much faster Variational Bayes method for approximating the posterior distribution of the geometric skew normal model is proposed and its performance is discussed. 
\end{abstract}


\section{Introduction}
\label{sec:intro}

 Pronounced skewness in data will lead to erroneous inferences with standard statistical procedures, For example, a standard confidence interval for the mean will be inaccurate, that is, the true coverage level will differ from the nominal level. Hence, the need to model skewed data has led to the development of many skewed distributions. Largely motivated by applications, interest in skewed distributions has grown immensely over the last few years. It is a fact that a majority of real data sets exhibit a deviation from symmetry rather than perfect symmetry (\cite{skeweg}).
 
 Likewise, skewed distributions have been used as alternatives in the construction of tests for symmetry and related inferential procedures(\cite{skew}). Several skewed distributions are obtained by adding a parameter that controls skewness to a symmetric distribution. The skew normal distribution proposed by \cite{ASN} is a popular example of a skewed distribution. A few other notable examples include the skew-t (\cite{skewt}), skew logistic (\cite{skewlogistic}) and skew cauchy (\cite{skewcauchy})  distributions. Finally, a generalized approach to generate a corresponding skew distribution from a symmetric distribution has also been proposed.(\cite{gskew})

The probability density function of the Azzalini skew-normal distribution (referred to as ASN) is as follows:
\[ f(x)=\frac{2}{\sigma}\phi(\frac{x-\mu}{\sigma})\Phi(\lambda\frac{x-\mu}{\sigma} )\qquad x \in \bm{R}
\]
where $\phi $ and $\Phi $ denote the standard normal PDF and the standard normal cumulative distribution function (CDF), respectively. It should be noted that the normal distribution becomes a particular member of this class of distributions ($\lambda=0$). It is unimodal density function and can have positive ($\lambda > 0$) as well as negative skewness($\lambda < 0$) . The skewness parameter, $\lambda$, also controls the shape of the distribution, thereby giving a flexible nature to its PDF. Furthermore, the ASN model has a natural multivariate generalization.(\cite{masn}). Unsurprisingly, this model has been implemented in analyzing non-symmetric data sets from various fields. (\cite{actuarial}, \cite{bio}, \cite{SPC})

Despite its numerous interesting properties, the ASN distribution suffers from several inferential problems as enumerated in the review paper by \cite{Review}. First, the method of moments estimator might not be in real values. Second , if $\mu=0$ and $\sigma=0 $ and all observations are positive (negative) the likelihood function will be monotonically increasing (decreasing) in $ \lambda $. Hence, the maximum likelihood estimate (MLE) of $ \lambda$ (minus) will be inﬁnite. Third, the MLE is not stable even when negative and positive observations are present. Fourth, there is the problem of identifiability with reference to the paraemter $ \lambda $, i.e., different values of $\lambda $ might correspond to very similar skew-normal densities. Next, analytically tractable standard errors and conﬁdence intervals cannot be obtained from the sampling distribution of the MLE. Lastly, if all three parameters are assumed to be unknown, the likelihood function for $ \lambda $ has a stationary point at $\lambda=0$, regardless of the observed sample. Finally, the Fisher information matrix becomes singular if $\lambda $ approaches zero.

The Bayesian approach can satisfactorily address both the problems of point estimation and hypothesis
testing of the skewness parameter. The work of \cite{reference} introduces a Jeffreys' reference prior for Bayesian inference and further, showing that the posterior has unbounded support and is proper, if the location and scale parameters are given and
known. \cite{Bayes} highlighted the advantages of the Bayesian approach with two different non-informative priors, the first being the Jeffrey's prior and the second an uniform prior. In the same paper, they also implement a Gibbs sampling algorithm for Bayesian analysis of several datasets including the frontier dataset, a dataset available on Azzalini's web page for which the classical MLE is infinite. 

The problem of prior elicitation for ASN distribution can be particularly challenging as $\lambda$ not only controls for symmetry but also influences the spread, modes and tail behaviour. \cite{Review} provide a comprehensive description of several prior elicitation schemes. For the goals of this article, the method prescribed by \cite{informative} using informative priors has been utilised.

Recently, another skew normal distribution based on a convolution of normal and geometric densities called the Geometric Skew Normal (referred to as GSN) distribution has been introduced by \cite{GSN}. The univariate GSN distribution is also three-parameter distribution and is a generalization of the normal distribution. In terms of skewness, the GSN also allows for a large degree of flexibility and more importantly does not suffer from the inferential issues of the ASN. \cite{GSN} has implemented and shown that a simple EM algorithm converges satisfactorily. The multivariate generalization of the GSN has also been proposed by \cite{mgsn}.

Surprisingly, regardless of the rising interest in skewed distributions, only a few papers deal with comparing the performance of different skewed distribution (\cite{comparison1}, \cite{comparison2} and \cite{comparison3}). Particularly, the the performance of the ASN distribution has not been compared with the recently proposed GSN for a variety of data sets. The aim of this article is to perform a thorough comparison of the ASN and GSN for modelling data sets with a wide range of skewness in the Bayesian paradigm, with the intention to benefit statistical modeling of skewed data sets.

First, to enable a beneficial and impartial comparison a Bayesian method of inference using informative priors for the parameters of the Geometric Skew Normal(GSN) is introduced. Next, the computation of the posterior using an MCMC method based on the Metropolis-Hastings algorithm is described. Alternatively, a Variational Bayes technique is also proposed for approximate Bayesian analysis of the posterior.  Finally, Bayesian inference with informative priors is implemented for the ASN distribution to compare the posterior predictive fit with that of the MCMC based Bayesian scheme for the GSN distribution for data sets with varying skewness. 
The paper is organized as follows: Section 2 describes the GSN distribution and discusses its properties. Section 3 provides the Bayesian scheme as well as the prior assumptions. Section 4 deals with comparison study of the ASN and GSN distributions using simulated and real data sets. Section 5 describes the alternate, approximate scheme of Bayesian inference for the GSN distribution using variational inference. Finally, in section 6 we conclude the paper and suggest future avenues of research based on the results we have obtained.
\section{Geometric Skew Normal Distribution (GSN)}
\label{sec:clarExpos}

In this section, we describe the Geometric skew-normal (GSN) proposed by \cite{GSN}. It is a skewed version of the normal distribution, in the sense that normal distribution is a particular member of this family. GSN enjoys several noteworthy properties. For example, the density of GSN distribution can be unimodal or multimodal, explicit forms of the moment generating function exist, and  GSN distribution can also have a symmetric distribution with heavier tails. The GSN distribution is characterized with three parameters $\mu$, $\sigma^2$ and p. To be precise, we provide the definition of GSN below.

Let $N$ and $X_i$, with $i=1,2,\ldots$, be independent random variables with 
    $N\sim \mbox{GE}(p) \mbox{ and } X_i\sim N(\mu,\sigma^2),$ where GE stands for geometric distribution. 
Define \[X=\sum_{i=1}^{N}X_i.\]
$X$ is said to have Geometric Skew Normal distribution and is denoted by  $X\sim GSN(\mu,\sigma,p).$

The distribution reduces to the normal density when p=1. The joint probability density function of $X$ and $N$, for $p \neq 1$, is given as
\[ f_{X,N}(x,n)= \phi((x-n\mu)/\sqrt{n}\sigma)p(1-p)^{n-1}/\sqrt{n}\sigma,
\]
where $\phi(\cdot)$ is the density function of the standard normal distribution. 
From the joint density of $X$ and $N$, the marginal density of $X$ can be obtained as
\begin{align}
\label{pdf: GSN}
f_X(x)=\sum_{k=1}^{\infty}\phi((x-k\mu)/\sqrt{k}\sigma)p(1-p)^{k-1}/\sqrt{k}\sigma.
\end{align}
Conditional density of $N$ given $X=x$ will be of our interest in future. In particular, the conditional expectations of $N$ and $N^{-1}$ are derived as follows (see \cite{GSN}):
\begin{align}
\label{expectation of N}
E(N \mid X=x)=\frac{\sum^{\infty}_{n=1}(1-p)^{n-1}e^{-\frac{1}{\sigma^2n}(x-n\mu)^2}.\sqrt{n}}{\sum_{k=1}^{\infty}(1-p)^{k-1}e^{-\frac{1}{\sigma^2k}(x-k\mu)^2}/\sqrt{k}}
\end{align}
and 
\begin{align}
\label{expectation of N-inv}
E(N ^{-1} \mid X=x)=\frac{\sum^{\infty}_{n=1}(1-p)^{n-1}e^{-\frac{1}{\sigma^2n}(x-n\mu)^2}/n^{3/2}}{\sum_{k=1}^{\infty}(1-p)^{k-1}e^{-\frac{1}{\sigma^2k}(x-k\mu)^2}/\sqrt{k}}.
\end{align}

\section{Bayesian Inference for Geometric Skew Normal}
\label{sec:titPage}
For performing Bayesian analysis, we first need to specify the prior distributions on the parameters of interests. 
The next subsection provides the details of the prior distributions. 
\subsection{Prior Assumptions}
\label{priors}
We consider the following informative conjugate priors on $\mu$, $\sigma$ , p  as follows:
\begin{align*}
   (\mu,\sigma^2) & \sim N-\Gamma(v_0,n_0,\alpha,\beta), \text{ and }\\
    p & \sim \mbox{Beta}(a,b),
\end{align*}
where $N-\Gamma$ denotes the Normal-Inverse Gamma distribution.
The choice of hyperparameters are discussed in subsections \ref{simu study} and  \ref{real data study} because the choices are made to be data dependent. 
\subsection{Posterior Analysis and Bayesian Framework}
Let $X_1, X_2, \ldots, X_{m}$ be iid random variables from a GSN with density provided in (\ref{pdf: GSN}). Then the likelihood is given as 
\begin{align}
\label{liklihood}
    L(\mu,\sigma,p) = \prod_{i=1}^m \sum_{k=1}^{\infty}\phi(x_i-k\mu/\sqrt{k}\sigma)p(1-p)^{k-1}/\sqrt{k}\sigma.
\end{align}
Using the the prior assumptions on $p$ and $(\mu,\sigma^2)$, specified in subsection~\ref{priors}, along with the likelihood given by (\ref{liklihood}) we obtain the full posterior distribution, $\pi (\mu,\sigma,p)$ as follows:
\[
\pi (\mu,\sigma,p\vert \mathbf{x}) \propto \prod_{i=1}^m \sum_{k=1}^{\infty}\phi(x_i-k\mu/\sqrt{k}\sigma)p^{a}(1-p)^{k+b-2}\frac{1}{\sqrt{k}}\left(\frac{1}{\sigma^2}\right)^{\alpha+2}e^{-\frac{2\beta+n_0(\mu-v_0)^2}{2\sigma^2}},
\]
where $\mathbf{x} = \{x_1, x_2, \ldots, x_{m}\}$ is the realization of $\mathbf{X} = \{X_1, X_2, \ldots, X_{m}\}$. However, the posterior density is not tractable because the posterior involves summation and therefore, even the full conditional densities of the parameters are not tractable. Instead, we consider the following hierarchical representation of the GSN random variables with latent variables $N_i$ corresponding to each $X_i$ for every $i=1,2, \ldots, m,$ thereby enabling a straight forward calculation of the conditional posteriors for our choice of conjugate priors. 
The hierarchical representation is given as follows
\begin{align}
    N_i\mid p & \sim GE(p). \label{N given p}\\
    X_i \mid N_i \, , \mu \, , \sigma \, , p & \sim N(N_i \mu,N_i \sigma^2) \text{ for } i=1,2, \ldots, m. \label{X given others}
\end{align}
%
With this representation, one can obtain the full conditional densities of the parameters, which facilitates the simulation from the joint posterior through Gibbs sampling. 

Let $\mathbf{N}$ denote the random vector $\{N_1, N_2, \ldots, N_{m}\}$ and let the realisation of $N_i$ be denoted as $n_i$, for $i=1,2, \ldots, m$. Then, using our choice of priors, the full conditional densities of $p$, and $(\mu,\sigma^2)$ are obtained as 
\begin{align}
p \mid \mu, \sigma,X, \mathbf{N}&\sim \mbox{Beta}(m+a,\sum_{i=1}^m n_i -m+b)
\label{full conditional of p} \\ 
(\mu,\sigma^2) \mid p,X,\mathbf{N} & \sim N-\Gamma(\mu^*,n^*,\alpha^*,\beta^*),
\label{full conditional of mu-sigma}
\end{align}
where
\begin{align*}
 & \mu^*=\frac{n_0v_0 + \sum_{i=1}^m x_i}{\sum_{i=1}^m n_i +n_0}, 
n^*=\sum_{i=1}^m n_i +n_0, 
\alpha^*= \alpha+ \frac{n}{2},  \text{ and }\\
&\beta^*= 2\beta +  n_0(u^*-v_0)^2 + \sum^m_{i=1}\frac{1}{n_i}(x_i-n_i\mu^*)^2. 
\end{align*}
The derivation of the full conditional densities are provided in the Appendix \ref{derivation of full conditional}. 

Since $\mathbf{N}$ is unobserved, we need an additional step to build a full sampling algorithm to obtain the joint posterior distribution of all unknown quantities in the model. For every  $i=1,2, \ldots, m$, corresponding to $X_i$, there is a latent variable $N_i$ whose posterior distribution depends only on $X_i$ and the other parameters of the model. The posterior distribution of $N_i$  for  $i=1,2, \ldots, m$, can be derived, up to a normalization constant, as follows:
\begin{align}
\label{full conditional of N}
    p(N_i \mid \mu, \sigma,p,X) \propto \frac{e^{-\frac{(X_i-N_i\mu)^2}{2N_i\sigma^2}}}{\sqrt{N_i}}(1-p)^{N_i-1}. 
\end{align}
%
\subsection{Hybrid Gibbs-Metropolis Hastings Algorithm}
In this section, we describe two hybrid Gibbs-Metropolis Hastings algorithms for simulation from the posteriors described in the last subsection. 
First, we update each $N_i$ for  $i=1,2, \ldots, m$ with a Metropolis Hastings algorithm using a geometric proposal. A similar method was implemented for Bayesian  Clustering using Multivariate GSN by \cite{clustering}.
Second, we also propose a random walk algorithm for updating each  $N_i$.
\subsubsection{Gibbs-Metropolitan Hastings Algorithm with Geometric Proposal}
\noindent
At the iteration $t$
        \begin{enumerate}
            \item  Sample $N^{(t)}$ (dependent on   $\sigma^{2^{(t-1)}}$ $\mu^{(t-1)}$ and $p^{(t-1)}$)
                 \begin{itemize}
                    \item Sample $p_{prop}$ from Uniform(0,1)
                    \item For each $i=1,2,\ldots,m$
                        \begin{itemize}
                            \item  Sample proposed $N_i^*$ from Geometric($p_{prop}$)
                            \item  Calculate the acceptance probability of $N_i$ using equation~(\ref{full conditional of N}) as
                            \[ r= \frac{p(N_i^* \mid X_i, \mu^{(t-1)}, \sigma^{2^{(t-1)}},p^{(t-1)})q(N_i^{(t-1)})}{p(N_i^{(t-1)}\mid X_i , \mu^{(t-1)}, \sigma^{2^{(t-1)}},p^{(t-1)})q(N_i^*)},
                            \]
                            where q is Geometric($p_{prop}$).
                            \item  Accept $N_i^*$ as the new sample $N_i^{t}$ with probability $\min(1,r)$.
                        \end{itemize}
                \end{itemize}
            \item Sample $\sigma^{2^{(t)}}$ (dependent on $N_i^{(t)}$), $\mu^{(t)}$ (dependent on $N_i^{(t)}$ and  $\sigma^{2^{(t)}}$ ), $p^{(t)}$ (dependent on $N_i^{(t)}$), from their respective conditional posteriors given in equations (\ref{full conditional of p}) and (\ref{full conditional of mu-sigma}). 
        \end{enumerate}

\subsubsection{Gibbs-Random Walk Algorithm}
\noindent
At the iteration $t$
        \begin{enumerate}
            \item Sample $N^{(t)}$ (dependent on  $\sigma^{2^{(t-1)}}$, $\mu^{(t-1)}$ and $p^{(t-1)}$)
                 \begin{itemize}
                    \item  For each i=1,2,...,m
                        \begin{itemize}
                            \item   Sample  $\epsilon$ with P($\epsilon \,$= 1) = 1/2 and P($\epsilon \,$= -1) = 1/2
                            \item   Simulate $N_i^*$ as: $ N_i^*=N_i + \epsilon$ (with constraint $N_i^* \geq 1$). That is, a random walk with reflecting boundary at 1. 
                            \item Calculate the acceptance probability of $N_i^*$, given in equation~(\ref{full conditional of N}) as
                            \[ r= \frac{p(N_i^* \mid X_i , \mu^{(t-1)}, \sigma^{2^{(t-1)}},p^{(t-1)})}{p(N_i^{(t-1)} \mid X_i , \mu^{(t-1)}, \sigma^{2^{(t-1)}},p^{(t-1)})}
                            \]
                            \item Accept $N_i^*$ as the new sample $N_i^{t}$ with probability $\min(1,r)$
                        \end{itemize}
                \end{itemize}
            \item Sample $\sigma^{2^{(t)}}$ (dependent on $N^{(t)}$),  $\mu^{(t)}$ (dependent on $N^{(t)}$ and  $\sigma^{2^{(t)}}$ ),  $p^{(t)}$ (dependent on $N^{(t)}$), from their respective conditional posteriors given in equations (\ref{full conditional of p}) and (\ref{full conditional of mu-sigma}).
        \end{enumerate}
 \section{Comparison study}
\label{comp study}

\subsection{Simulation Studies}
\label{simu study}
To compare the performance of the ASN and GSN in terms of fitting to a skewed data, we simulated from a varied range of skewness starting from very small skewness to large skewness. To make a fare comparison between ASN and GSN, majority of these data sets are either simulated from ASN. The data with very small skewness, small skewness and large skewness were generated from ASN. One data set with moderate skewness is simulated from GSN for illustration purpose. 
For generating ASN samples we used the R package `sn'. The samples from GSN were generated using equation (\ref{N given p}) and (\ref{X given others}) by:
%
\begin{itemize}
    \item Generate N from Geometric(p)
    \item Generate X from Normal($N\mu$,$N\sigma$). Consequently,  X follows GSN($\mu,\sigma$,p) 
\end{itemize}
We obtain the posterior predictive fit for GSN and ASN and compare the fit using the Kolmogorov-Smirnov distance (KSD) between the fit and the original data. We also report the maximum a posteriori (MAP) estimates of the parameters. The R function 'map\_estimate' was used to obtain the MAP. 

A final simulation study was done with data generated from a skewed distribution different from ASN and GSN. Particular, we have simulated from lognormal distribution and fitted the data with ASN and GSN distributions under Bayesian paradigm. In this case also, we report the KSD to compare the goodness of fit between ASN and GSN. 

In all these simulation studies, 100 samples are drawn from the respective distributions. 

For the doing the inference on the parameters of GSN, we use the Hybrid Metropolis-Gibbs algorithm with a random walk step for updating the latent variables. We note that using a geometric proposal distribution (\cite{clustering}) instead of a simple random walk produces similar results. 

We obtain the posterior predictive fit of ASN using the method of \cite{informative}. A brief description of the Gibbs Sampling algorithm has been provided in the appendix 

The prior elicitation for both models was done to reduce the influence of the prior distribution. By setting the hyper parameters accordingly we obtained nearly non-informative priors. Prior for the parameter $p$ in GSN model was set to Beta(1,1) which lends a non-informative uniform distribution. A diffuse normal prior was elicited for the mean by inflating the variance (setting hyperparamter $n_0=0.001$). Similarly for ASN, diffuse normal priors were elicited for the mean and skewness parameter $\lambda$.

\subsubsection{Very Small Skewness}
\begin{figure}[!h]
    \includegraphics[width=0.5\textwidth]{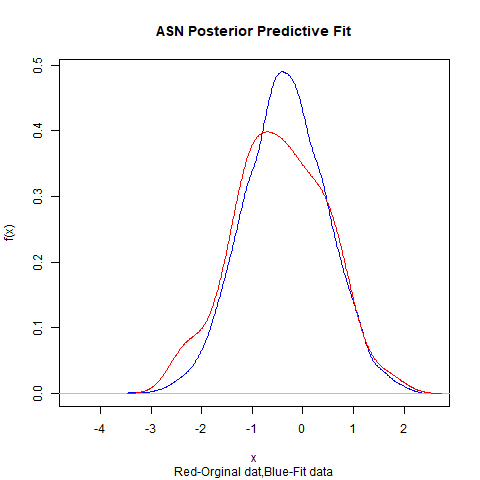}
    \includegraphics[width=0.5\textwidth]{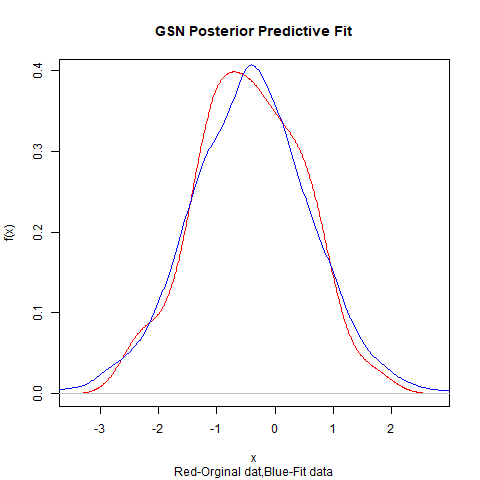}
    \caption{Very Small Skewness Data: In red is the kernel density of 100 smaples for ASN(($\mu$= 0,$\sigma$=1, $\lambda$=-0.5) and the in blue, the density of posterior predictive fit of the ASN and GSN models on the left and right respectively. A Bayesian scheme of inference with informative priors was used in both cases. }
    \label{fig 1: very small skewness}
\end{figure}

In this subsection, data are generated from ASN ($\mu=0, \sigma=1, \lambda = -0.5)$ so that the skewness is very small. Indeed, the generated sample had Pearson co-efficient of skewness -0.038. The results of the posterior predictive fit for both models is provided in Table \ref{table1: very small skewness}. Since the data were generated from ASN so we can compare the MAP estimate of parameters with the true values.  Table \ref{table1: very small skewness} shows that MAPs of the parameters of ASN distribution are very close to the true values. It is also seen that the length of the credible intervals for the parameters are small (for breivity the plots are not reported). The posterior predictive densities on top of the observed data density for ASN and GSN are depicted in the Figure \ref{fig 1: very small skewness}. The figure shows that the fit by GSN is better than that of ASN, which is also confirmed by the KSD statistics provided in the last column of Table~\ref{table1: very small skewness}. 
\begin{table}[!h]
\centering
\begin{tabular}{|l|r|r|r|r|} \hline
Model & Mean($\mu$) & Standard Deviation ($\sigma$) & $\lambda$(ASN) or p(GSN)  & KSD      \\ \hline
  ASN & 0.01  & 0.89 &  -0.5 &  0.119  \\ \hline
  GSN  &  -0.41 & 0.97 & 0.93 & 0.049 \\ 
   \hline
\end{tabular}
\caption{
Very Small Skewness Data: MAP values of the parameters of ASN and GSN are reported. Also the KSD between the original data and the posterior predictive fit for ASN and GSN. KSD for GSN is slightly better.}
\label{table1: very small skewness}
\end{table}

\vspace{-0.75cm}

\subsubsection{Small Skewness}
As mentioned in the beginning of the Section~\ref{simu study}, data are simulated from ASN which caters small skewness. The parameters for ASN are chosen to be $\mu= 0$, $\sigma=1$, $\lambda=-1$. The Pearson co-efficient of skewness turned out to be $-0.370$, which indicates a negative skewness as expected. Notice that the magnitude of the skewness is larger than the previous simulated data. 

Here also, we obtained the MAP of the parameters involved in ASN and GSN, respectively. Since the simulated data are from ASN, so the comparison with the MAP of the ASN parameters can be made. Table \ref{table2: small skewness} provides the MAPs of the parameters. Moreover, from the posterior density of the $\mu$ and $\sigma$ for ASN, it is observed that credible intervals are of small length. 
Although it is seen that the MAP of ASN parameters estimates the true parameter values quite well and the corresponding length of the credible intervals are small, the KSD of GSN turns out to be slightly smaller than that of ASN. 
\begin{table}[!h]
\centering
\begin{tabular}{|l|r|r|r|r|} \hline
Model & Mean($\mu$) & Standard Deviation ($\sigma$) & $\lambda$(ASN) or p(GSN)  & KSD      \\ \hline
  ASN &  0 & 0.98 & -1.22 &  0.084 \\ \hline
  GSN  & 0.58 & 0.89 & 0.86 & 0.059 \\ 
   \hline
\end{tabular}
\caption{Small Skewness Data: MAP values of the parameters of ASN and GSN are reported. Also the KSD between the original data and the posterior predictive fit for ASN and GSN.}
\label{table2: small skewness}
\end{table}
\begin{figure}[!h]
    \includegraphics[width=0.5\textwidth]{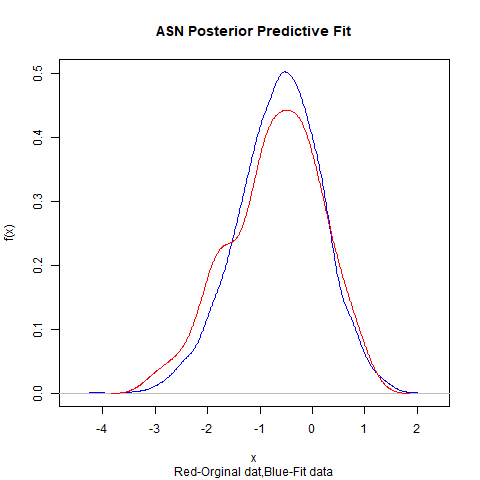}
    \includegraphics[width=0.5\textwidth]{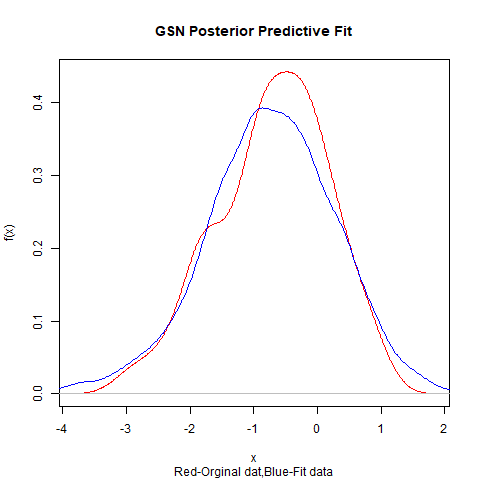}
    \caption{Small Skewness Data: In red is the kernel density of 100 smaples for ASN(($\mu$= 0,$\sigma$=1, $\lambda$=-1) in blue, the density of posterior predictive fit of the ASN and GSN models on the left and right respectively. A Bayesian scheme of inference with informative priors was used in both cases.}
    \label{fig 2: small skewness}
\end{figure}

\subsubsection{Moderate Skewness}
To generate samples with moderate skewness we take the help from GSN. One Hundred data points were sampled from GSN with the model parameters $\mu= 1$,$\sigma=1$ and $p=0.8$. The Pearson co-efficient of skewness of simulated data is found to be $0.748$, which is larger (in terms marginitude) than the previous two simulated data sets. 

\begin{figure}[!h]
    \includegraphics[width=0.5\textwidth]{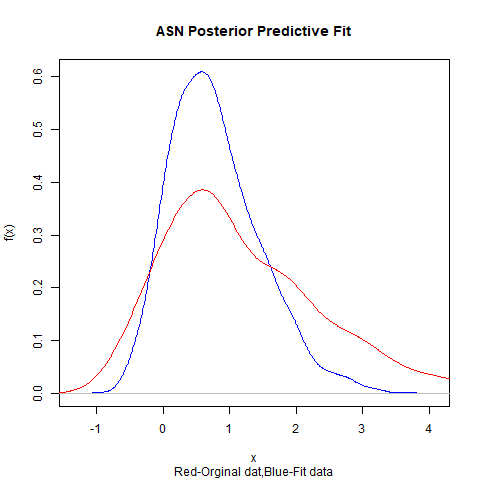}
    \includegraphics[width=0.5\textwidth]{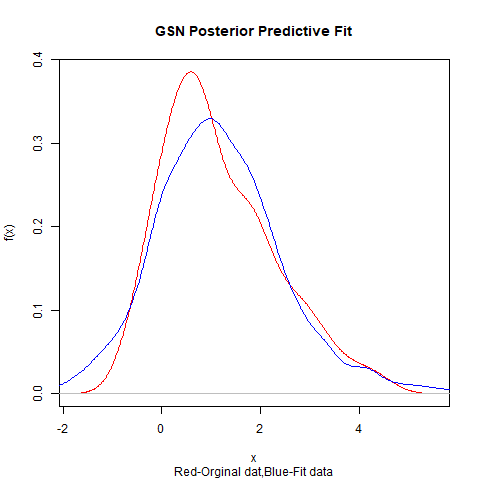}
    \caption{Moderate Skewness Data: In red is the kernel density of 100 samples for GSN(($\mu$=1,$\sigma$=1, $\lambda$=0.8) in blue, the density of posterior predictive fit of the ASN and GSN models on the left and right respectively. A Bayesian scheme of inference with informative priors was used in both cases.}
    \label{fig 3: moderate skewness}
\end{figure}

Now in this simulation study, the MAP estimates of the parameters of GSN can be compared with the true parameter values of GSN. Clearly (see Table \ref{Table3: Moderate skewness}) MAP of the GSN model provides excellent estimates of the true parameters of the simulated data. Although not reported here, the credible intervals for the parameters turns out to be very reasonable in terms of the length of the interval. 
Importantly, the KSD corresponding to the GSN model is considerably lower than that of the ASN model. This might not be surprising because the data were simulated from GSN itself. However, it is to be noticed that even when the data sets were generated from ASN the KSD corresponding to GSN were comparable to ASN (indeed, they were lower than the ASN). 
Figure \ref{fig 3: moderate skewness} depicts the fit of ASN and GSN to the original data. It is clear that GSN model provides a better approximation to the true data, especially near the right tail. In both the tails, ASN underestimates the true density.  
\begin{table}[!h]
\centering
\begin{tabular}{|l|r|r|r|r|} \hline
Model & Mean($\mu$) & Standard Deviation ($\sigma$) & $\lambda$(ASN) or p(GSN)  & KSD      \\ \hline
  ASN &  0 & 1.01 &  2.88 &  0.223 \\ \hline
  GSN  &  0.94 & 1.06 & 0.83  & 0.09 \\ 
   \hline
\end{tabular}
\caption{
Moderate Skewness Data: MAP values of the parameters of ASN and GSN are reported. Also the KSD between the original data and the posterior predictive fit for ASN and GSN.}
\label{Table3: Moderate skewness}
\end{table}

\subsubsection{Large Skewness}
Heavily positive skewed data are simulated with a very high value of $\lambda$ from ASN. We generate 100 random observations from ASN with $\mu$= 0,$\sigma$=1, $\lambda$=100. The Pearson co-efficient of skewness (0.956) of the simulated data indicated a large positive skewness compared to the previously simulated data sets. The MAP estimates of the parameters for ASN and GSN are shown in Table \ref{table 4: large skewness}. KSD values are also reported in the table \ref{table 4: large skewness}. 

\begin{figure}[!h]
    \includegraphics[width=0.5\textwidth]{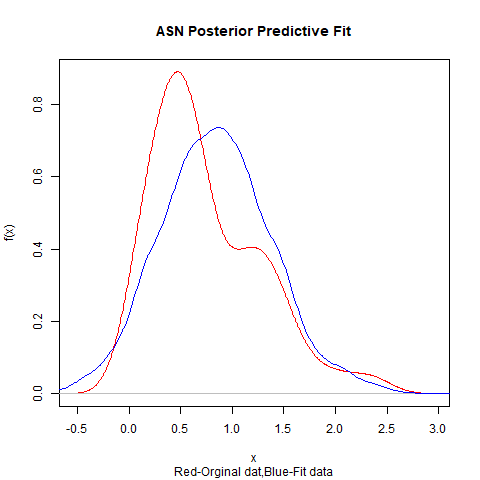}
    \includegraphics[width=0.5\textwidth]{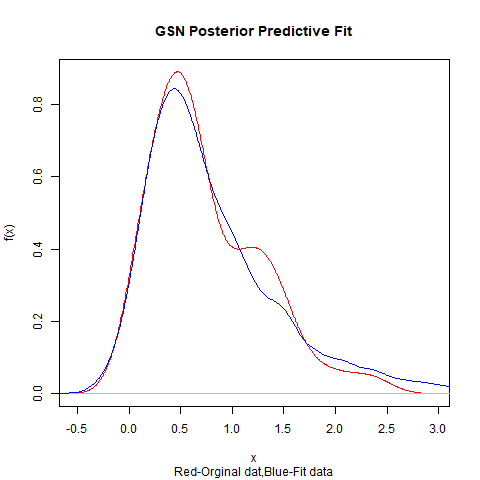}
    \caption{Large Skewness Data: In red is the kernel density of 100 smaples for ASN(($\mu$= 0,$\sigma$=1, $\lambda$=100) in blue, the density of posterior predictive fit of the ASN and GSN models on the left and right respectively. A Bayesian scheme of inference with informative priors was used in both cases.}
    \label{fig 4: large skewness}
\end{figure}

A few important things to be noted, as the skewness becomes as high as 0.956, the MAP of ASN comes out to be poor. In fact, the 95\% credible interval fails to capture the true values, for both $\mu$ and $\sigma$. Besides, the KSD for ASN is much larger than the KSD for GSN. Figure~\ref{fig 4: large skewness} also confirms the finding. The fitted density for ASN does a poor job compared to GSN. Clearly, for this data set, GSN outperforms the ASN. 

\begin{table}[!h]
\centering
\begin{tabular}{|l|r|r|r|r|} \hline
Model & Mean($\mu$) & Standard Deviation ($\sigma$) & $\lambda$(ASN) or p(GSN)  & KSD      \\ \hline
  ASN & 0.48 & 0.64 &  0.94 &  0.207 \\ \hline
  GSN  &  0.36 & 0.26 & 0.44 & 0.07 \\ 
   \hline
\end{tabular}
\caption{
Large Skewness Data: MAP values of the parameters of ASN and GSN are reported. Also the KSD between the original data and the posterior predictive fit for ASN and GSN. }
\label{table 4: large skewness}
\end{table}

\subsubsection{Lognormal}

\begin{figure}[!h]
    \includegraphics[width=0.5\textwidth]{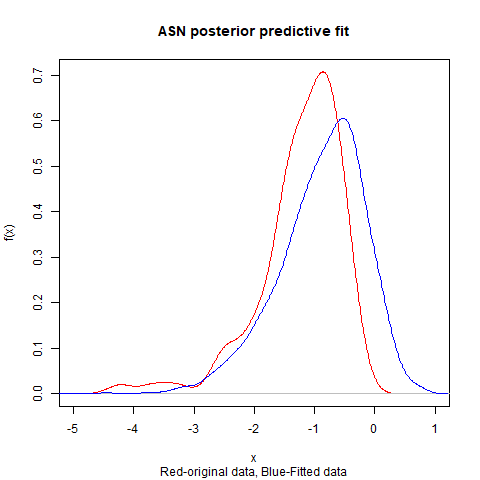}
    \includegraphics[width=0.5\textwidth]{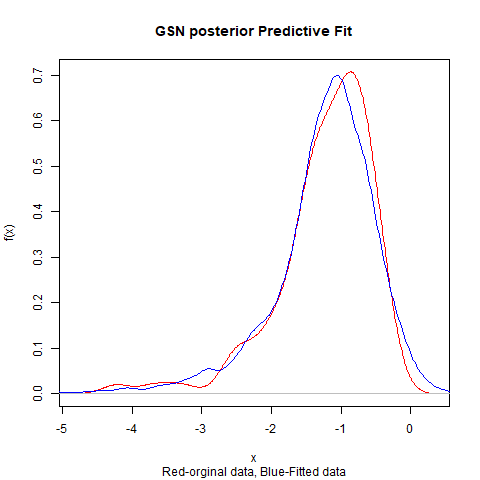}
    \caption{Lognormal Data: In red is the kernel density of 100 smaples from a Log normal density. The sample has a large negative co-efficient of skewness(-1.641). In blue, the density of posterior predictive fit of the ASN and GSN models on the left and right respectively. A Bayesian scheme of inference with informative priors was used in both cases.}
    \label{fig 5: lognormal}
\end{figure}

As a final simulation study, we simulate 100 random observations from a log-normal density with paraemters $\mu$=0 and $\sigma^2=0.6$. The negative values of the generated sample was considered, so as to obtain a negative Pearson co-efficient skewness of -1.641. Hence this particular data set has the largest skewness (in magnitude) amongst all the data sets considered here. In this case, we provide the plots of the true density and the fitted density for ASN and GSN (Figure~\ref{fig 5: lognormal}). The fitted density of GSN model describes the data better than the fitted density of ASN. KSD between GSN posterior predictive and the original sample is 0.063 as against 0.278 in the case of ASN. In this highly negative skewed data as well, GSN performs significantly better than ASN. 

\subsubsection{Findings of the Simulation study}
\label{conclusion of simu study}

From the results we note the following:
\begin{enumerate}
    \item  First, for data sets with small deviations from normality, that is for datsets with small coefficients of skewness the fit of ASN and GSN are comparable. 
    \item Data with larger tails can be captured better by GSN.
    \item In the presence of moderate skewness we see that GSN once again works better.
    \item When the skewness of the dataset is large, the GSN model clearly out performs ASN. This fact is hardly surprising since the ASN distribution only allows for a maximum coefficient of skewness of 0.995, in the case when the skewness parameter $\lambda= \pm \infty$
\end{enumerate}

\subsection{Real Data Analysis}
\label{real data study}
In this section, the performance of ASN and GSN are compared using two real data sets. 
\subsubsection{Frontier Data Set}
\begin{figure}[!h]
    \includegraphics[width=0.5\textwidth]{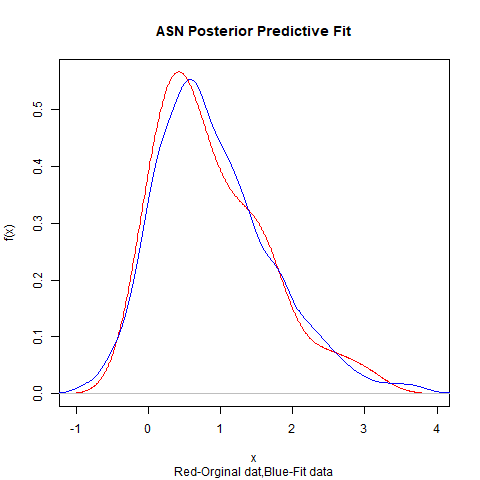}
    \includegraphics[width=0.5\textwidth]{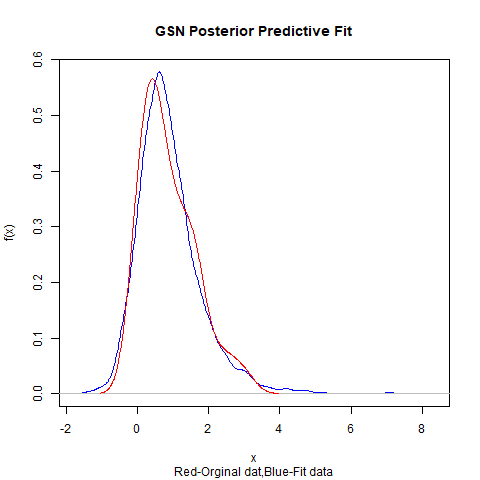}
    \caption{In red is the kernel density of the frontier data set from Azzalini's webpage. In blue, the density of posterior predictive fit of the ASN and GSN models on the left and right respectively. A Bayesian scheme of inference with informative priors was used in both cases. }
    \label{fig 6: frontier data}
\end{figure}

First we analyze Frontier data set for comparison of two skew-normal distributions. 
`Frontier data set' is available on the webpage of Adelchi Azzalini \url{http://azzalini.stat.unipd.it/index-en.html}. The classical MLE of the skewness parameter  for this data set is $\infty$, thereby rendering a half normal density, however it can be easily seen that finite values of the skewness parameter give a far better fit. \cite{Bayes} used this data set to call attention to the benefits of the Bayesian scheme. For this data set, we once again report the KSD between the ASN and GSN posterior predictive fit (Figure~\ref{fig 6: frontier data}) and the original sample as 0.096 and 0.089, respectively. This reconfirms the conclusions of \cite{Bayes}, that the Bayesian method can work for data where the classical MLE fails and also that the GSN model can be used as an alternative to the ASN.

\subsubsection{Guinea Pig data}

Finally, we consider a real data set representing the survival times of guinea pigs injected with different doses of tubercle bacilli. This data set has been obtained from \cite{GSN}, where the author uses this data set to illustrate the advantage of using GSN over ASN for classical MLE inference. The data set has a fairly large Pearson co-efficient of skewness of 1.796. Consequently, from fig 7 it is evident that GSN model provides a better fit than the ASN model. Moreover we have the KSD values 0.313 and 0.12 for the ASN and GSN posterior predictive fits respectively. Interestingly, we would like to point out that in the paper \cite{GSN}, the author obtains the corresponding KSD, between the fitted distribution using MLE estimates and the original density of the data to be 0.13. Hence, we can conclude that the our Bayesian scheme for GSN works marginally better than classical MLE inference. Nevertheless, for this data set the GSN model works better than the ASN model, in both the classical and Bayesian schemes.

\begin{figure}[!h]
    \includegraphics[width=0.5\textwidth]{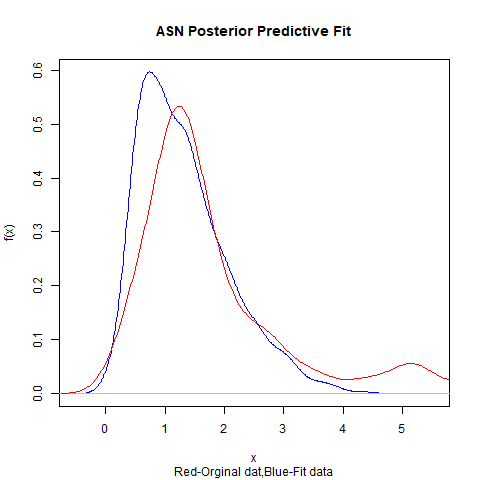}
    \includegraphics[width=0.5\textwidth]{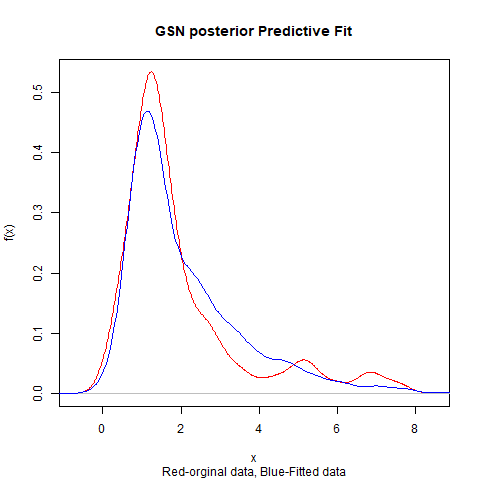}
    \caption{In red is the kernel density of the guinea pig dataset from \cite{GSN}. The sample has a large positive co-efficient of skewness(1.796). In blue, the density of posterior predictive fit of the ASN and GSN models on the left and right respectively. A Bayesian scheme of inference with informative priors was used in both cases.}
    \label{fig 7: guinea pig data}
\end{figure}
\section{Variational Inference for Geometric Skewed Normal}
\label{sec:noteConv}

\subsection{Primer on Variational Inference (VI)}

Variational inference (VI) has been widely used to approximate posterior densities for Bayesian models. In comparison to MCMC, VI tends to be much faster and can be easily scaled to large data sets. It has been applied to problems such as large-scale document analysis, computational neuroscience, and computer vision. However, unlike MCMC, VI does not have a guarantee of asymptotically exact convergence to the posterior.  We describe the preliminaries of VI as given in \cite{VI}. 

Let $\bm{x} = \{x_1,x_2,...x_n\}$ be a set of observed variables and $\bm{z} = \{z_1,z_2,....z_m\}$ be a set of latent variables, with joint
density $p(\bm{z}, \bm{x})$. We assume that all unknown quantities of interest are represented as latent random variables. This includes parameters that might determine the data, and latent variables associated with each data point. The goal of the inference problem is to compute the conditional density of the latent variables given the observations, $p(\bm{z} \mid \bm{x})$, which can be written as
\[p(\bm{z} \mid \bm{x})= \frac{ p(\bm{z}, \bm{x})}{p(\bm{x})}.
\]
The denominator is the marginal density or the model evidence, provided as 
\[ p(\bm{x})= \int p(\bm{z}, \bm{x}) dz. 
\]
For many models, this evidence integral is unavailable in closed form or is computationally impractical. Inference in such models is hard because we require the evidence to compute the conditional from the joint.

%
In VI, a family $\mathcal{Q}$ of densities over the latent variables is specified. Each element of $\mathcal{Q}$, say $q(\bm{z})$, is a candidate density for approximation to the exact conditional. Our goal is to find the
best candidate, that is, the one closest in KL divergence to the exact conditional. Inference is now reduced to solving the optimization problem:
\[q^*(\bm{z})= \argmin_{q(\bm{z}) \epsilon \mathcal{Q}} KL(q(\bm{z}) \mid \mid p(\bm{z} \mid \bm{x})), \]
where the KL divergence is expressed as:
\[
KL (q(\bm{z}) \mid \mid p(\bm{z} \mid \bm{x})) = E[\log q(\bm{z})]-E[\log p(\bm{z} \mid \bm{x})].
\]
Note that all expectations are taken with respect to $q(\bm{z})$.
By expanding the conditional, we get 
\begin{align}
\label{KL}
KL (q(\bm{z})\mid \mid p(\bm{z}\mid \bm{x})) = E[\log q(\bm{z})] - E[\log p(\bm{z}, \bm{x})] + \log p(\bm{x}), 
\end{align}
which depends on log p(x) and hence, cannot be computed easily. To address this, instead of computing the KL, we maximize an alternative objective that is equivalent
to minimizing the KL up to an added constant. The alternative objective function is known as evidence lower bound (ELBO) and is given by: 
\begin{align}
\label{ELBO}
    \mbox{ELBO}(q) = E[\log p(\bm{z}, \bm{x})] - E [\log q(\bm{z})]. 
\end{align}
It is called evidence lower bound because ELBO gives a lower bound of the (log) evidence, namely, $\log p(x) \geq \mbox{ELBO}(q)$, 
for any $q(\bm{z})$. To observe this, notice that Equation (\ref{KL}) and (\ref{ELBO}) give the following expression of the evidence,
\[\log p(x) = KL (q(\bm{z})\mid \mid p(\bm{z} \mid \bm{x})) + ELBO(q),
\]
and thus, $\log(p(\bm{x})) \geq \mbox{ELBO}(q)$, for any $q\in \mathcal{Q}$. 
\subsection{Mean field Approximation}
If the latent variable can be grouped into independent components, then the calculation of ELBO can be simplified. To achieve this, we utilise the mean field approximation (Chapter ... of \cite{Bishop}). In particular, for GSN the idea of mean field approximation will be very convenient. 
Let the elements of $\bm{z}$ be partitioned into disjoint groups $z_i$ where $i=1,2 \ldots m$. The mean field approximation specifies the family of distributions  $\mathcal{Q}$ to those distribution $q(z)$ which factorize as follows:
    \[ q(\bm{z})= \prod_{i=1}^mq_i(z_i).
    \]
Amongst all distributions $q(\bm{z})$, which factorize, we seek that distribution for which the  ELBO$(q)$ is largest. Therefore, we wish to make an optimisation of ELBO{L}$(q)$ with respect to all the distributions $q_i(z_i)$. The algorithm we have used for solving this optimization problem is the coordinate ascent variational inference
(CAVI) (see \cite{Bishop} for details). CAVI iteratively optimizes each factor of the mean-field approxiamtion
density, while holding the others fixed. It ascends the ELBO to a local optimum.

According to \cite{Bishop}, the optimal solution, $q_j^*(z_j)$, is the expectation of $\log p(\bm{x},\bm{z})$ with respect to $q_i(z_i)$ for all $i \neq j$, that is, 
        \[ \log q_j^*(z_j)=E_{i \neq j}[\log p(\bm{x},\bm{z})] + \mbox{const}.
        \]     
We initialize all of the factors $q_i(z_i)$ ,$i=1,2,\ldots, ,m$, and then cycle through the factors and replace each in turn with a revised estimate given by the above equation, using the current estimates for all of the other factors. The detailed algorithm for GSN is given in the Section \ref{Algo and simu: variational bayes}. 
\subsection{Variational Inference for GSN}
From the PDF of GSN random variable $X$, we observe that GSN law can be viewed as a geometric mixture of normal random variables, i.e.,  
    \[ f_{X\mid (\mu,\sigma,p)}(x)=\sum_{k=1}^{\infty} \frac{p(1-p)^{k-1}}{\sqrt{k}\sigma}\phi(x-k\mu/\sqrt{k}\sigma). 
    \]
It has been observed that for mixture models, where Gibbs sampling is not an option, VI may perform better than a more general MCMC technique (\cite{VI}). This motivates us to explore a variational bayes technique for the GSN using the CAVI algorithm.

Similar to the MCMC algorithm, we assume the following informative priors on $\mu,\sigma^2,p$: 
\begin{align*}
p & \sim \mbox{Beta}(a,b), \mbox{ and } \\
(\mu,\sigma^2) & \sim N-\Gamma(v_0,n_0,\alpha,\beta), 
\end{align*}
where $N-\Gamma$ denotes the Normal-Inverse Gamma distribution.
Recall our model for GSN random variables $X_1,...X_n$:
\[N_i\mid p \sim Geometric(p)\]
\[X_i \mid N_i \, , \mu \, , \sigma \, , p \sim N(N_i \mu,N_i \sigma^2) \qquad i=1,2,...m\] 
According to the terminology of VI, let the vector of latent variables be $\bm{z}=(\bm{N},\mu,\sigma^2,p)$, with $\bm{N}= (N_1,N_2,...N_m)$, which includes both parameters and the unobserved variables. Assume that the variational distribution $q$ factorizes as :
    \[ q(\bm{z})=q(\bm{N})q(\mu,\sigma,p).
    \]
We obtain the following optimal variational distribution for the latent variables $\bm{N}$ as (for the full derivation see the appendix):
\begin{align}
\label{VI: N}
    q^*(\bm{N}) = \prod_{i=1}^m \frac{1}{\bm{C}_i\sqrt{n_i}}e^{An_i +B\frac{x_i^2}{n_i}},
\end{align}
where
\begin{align*}
      \bm{C}_i = \sum_{i=1}^\infty \frac{1}{\sqrt{n_i}}e^{An_i + B\frac{x_i^2}{n_i}}, \text{ with }
        A  =\psi(b^*)-\psi(a^*+b^*) -\frac{1}{2n^*}-\frac{(\mu^*)^2\alpha^*}{2\beta^*}, \text{ and }
     B =-\frac{\alpha^*}{2\beta^*}.
\end{align*}
Thanks to the conjugate prior choice, we obtain the following closed forms of the optimal variational distributions for the model parameters $(p,\mu,\sigma)$:
\begin{align}
\label{VI: p}
    q^*(p) \sim \mbox{Beta}(a^*, b^*), 
\end{align}   
\text{ where } $ a^*= n+a, \text {and }
                b^*= E_N\left(\sum_{i=1}^m n_i\right)-n+b.$ 
\begin{align}
\label{VI: mu-sigma}
     q^*(\mu , \sigma^2) \sim N-\Gamma(\mu^*,n^*, \alpha^*, \beta^*), 
\end{align}                
where 
\begin{align*}
\mu^*&=\frac{\sum_{i=1}^m x_i + n_0v_0}{n_0+E_N\left( \sum_{i=1}^m n_i \right)}, \,\, n^*=n_0 + E_N\left(\sum_{i=1}^m n_i\right), \,\, \alpha^*  =\alpha +1+\frac{n}{2}, \text{ and }\\
 \beta^*& =2 \beta  + n_0(\mu^*-v_0)^2 + E_N\left( \sum_{i=1}^m \frac{1}{n_i}(x_i-n_i\mu^*)^2\right). 
\end{align*}
Further we note that the expectations with respect to N can evaluated as follows using the expectations given in (\ref{expectation of N}) and (\ref{expectation of N-inv}):
\begin{align}
\label{eqns: expectations in VI}
     E_N\left(\sum_{i=1}^m n_i\right) & =  \sum_{i=1}^m E_{N_i}(n_i) \\
    E_N\left( \sum_{i=1}^m \frac{1}{n_i}(x_i-n_i\mu^*)^2\right) & = \sum_{i=1}^m \left( x_i^2E_{N_i}\left(\frac{1}{n_i}\right)+ \mu^{*2}E_{N_i}(n_i) - 2x_i\mu^*\right)
\end{align}
\subsection{Algorithm and Simulation}
\label{Algo and simu: variational bayes}
The CAVI algorithm for computing the Variational distribution, for GSN, as an approximation to the posterior is described below:

\begin{itemize}
    \item Initialize $N_i^{(0)},\mu^{(0)},{\sigma^2}^{(0)},p^{(0)}$ and compute $ELBO^{(0)}(q^{*}(N_i^{(0)},\mu^{(0)},{\sigma^2}^{(0)},p^{(0)})$
    \item For iteration t, while $ELBO^{(t-1)} - ELBO^{(t-2)}> \epsilon$
        \begin{enumerate}
            \item Compute the variational distribution $q^*(N^{(t)})$ given in equation (\ref{VI: N}) using  $\mu^{(t-1)},{\sigma^{2}}^{(t-1)},p^{(t-1)}$ and compute the required expectations  using equations (\ref{expectation of N}) and (\ref{expectation of N-inv}).
            \item Compute the variational distributions of  $\mu^{(t)},{\sigma^2}^{(t)},p^{(t)}$ from equations (\ref{VI: p}) and (\ref{VI: mu-sigma}) and re-compute the expectations calculated in the step 1. 
            \item Calculate $ELBO^{(t)}(q^*(N,\mu,\sigma,p))$ given by equation (\ref{ELBO}) (see appendix for the full expression of ELBO).
          \end{enumerate}
    \item return $q^*(\mu,\sigma,p)$
\end{itemize}

\begin{figure}[!h]
    \includegraphics[width=0.5\textwidth]{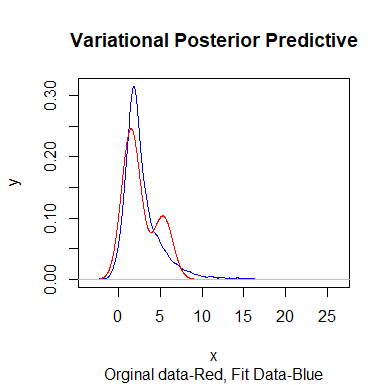}
    \includegraphics[width=0.5\textwidth]{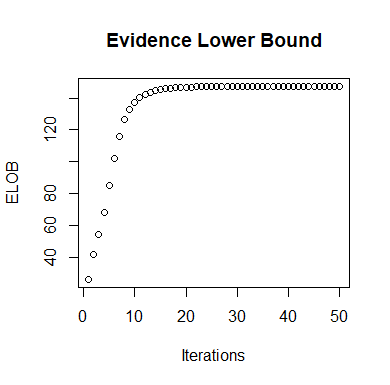}
    \caption{Left: Plot of the Variational Posterior Predictive Fit  (in blue) against the original 100 sample from $GSN(\mu=2,\sigma=1,p=0.6)$ (in red). The Variational posterior distribution was obtained using the CAVI algorithm Right: Plot of the ELBO for the GSN model. The ELBO increases with each iteration and converges quickly, showing that the CAVI algorithm optimizes the ELBO.}
    \label{fig 8: variational fit toy example}
\end{figure}

To illustrate the  convergence of the CAVI algorithm, we generated a random data set from $GSN(\mu=2,\sigma=1,p=0.6)$ and implemented the algorithm. The Variational posterior predictive as well as the convergence of the ELBO have been shown in Figure~\ref{fig 8: variational fit toy example}. 

\begin{table}[!h]
\centering
\begin{tabular}{|l|r|r|r|r|} \hline
Data Set & GSN- VI & GSN-MCMC  & ASN    \\ \hline
  Very Small Skewness &  0.125 & 0.049 & 0.119  \\ \hline
  Small Skewness  &  0.129  & 0.059 & 0.084 \\ \hline
  Moderate Skewness  &   0.131  & 0.09  & 0.223   \\ \hline
  Large Skewness  &  0.130 & 0.07  & 0.207\\ \hline
  Lognormal  &  0.148 & 0.063 &  0.278 \\ \hline
  Frontier Data &  0.164 & 0.089  & 0.096 \\ \hline
  Guinea Pig Data &  0.203 &  0.12 & 0.313  \\
  \hline
\end{tabular}
\caption{
Kolmogorov-Smirnov Distance(KSD) of the Varaitonal Posterior predictive and the original data along with corresponding  KSD results for ASN and GSN models for the data sets used in comparison study( Section \ref{simu study}) }
\label{Table: Variational Results}
\end{table}
To see the performance of VI comparing with the performance of MCMC technique on GSN and on ASN, we implemented our variational technique on the simulated data sets and real sets considered in Section~\ref{simu study} and \ref{real data study}. The KSD values are reported in Table ~\ref{Table: Variational Results}. For very small to small skewness data, ASN using MCMC technique has done better than variational mehods on GSN. However, as soon as the skewness increases to moderate or large skewness  data, GSN with variational Bayes, which is an approximate technique, performs better than ASN. Further, similar reults are seen in the case of the real data sets as well. However, through out all these studies, GSN based on MCMC works better than other two. It is not surprising that GSN with variational method will pefrorm poorly when compared to an MCMC technique, as the first method is an approximation method. However, That said, it has to be kept in mind the computation time is significantly improved for the variatioanal method over the MCMC methods. 
\section{Conclusion}
\label{sec:eqnNumb}
We have described the problem of handling skewness in data along with two skew normal distributions, the well known Azzalini Skew Normal(\cite{ASN}) and the recently proposed Geometric skew Normal (\cite{GSN}). We have noted the advantages of Bayesian Inference for Azzalini skew-normal and we further described Bayesian inference for GSN with informative priors using a hybrid-Metropolis Hastings algorihtm as well as proposed a novel Variational Bayes method. 

From the results of our simulation studies and real data analysis, we observe that for small to moderate skewness the fit of ASN and GSN are comparable. However, data with large skewness and heavy tails can be captured better by GSN. We also wish to note, implementation of Bayesian inference and selecting informative priors is easier in the case of GSN. Furthermore, a simple Variational algorithm is shown to converge quickly in the case of GSN.

ASN has an advantage with regards to small deviations from normality as skewness can be controlled with a single parameter, while in the case of GSN both the mean ($\mu$) and the probability ($p$) determine skewness. However as mentioned in the introduction, the prior elicitation of the skewness parameter($\lambda$) for ASN can be difficult since $\lambda$ controls not only symmetry but also the shape of the distribution. Furthermore, implementation of informative Bayesian can be more challenging, as the conditional posteriors belong to the Unified family of Skew-Normal distribution.(\cite{informative}).

Finally, we have mentioned that multivariate extensions of both GSN (\cite{mgsn}) and ASN(\cite{masn}) exist, Bayesian inference can be further extended for both and similarly compared. A mixture distribution of multivariate GSN or ASN distributions can be applied in clustering probelms (\cite{clustering}) . GSN can be applied to Bayesian regression to compare the performance of GSN against ASN for skew normal variates. The Variational Inference can motivate the development of faster and easily scalable methodology for Bayesian regression in high dimensions (\cite{vskew}).
\section{Acknowledgements}
\label{sec:prepDocs}

I would like to thank my supervisor Dr. Satyaki Mazumder for his insight on the topic of skew normal and his constant support throughout this project up and until the writing and editing of this paper.
\section{Appendix}
\label{sec:concComm}

\subsection{Derivation of Conditional Posterior Distributions}
\label{derivation of full conditional}
Once again let $\mathbf{X} = \{X_1, X_2, \ldots, X_{m}\}$ be the GSN random variables and let  $\mathbf{N}=\{N_1, N_2, \ldots, N_{m}\}$ be the latent variables.
Considering the following prior for p:
\[ p \sim \text{Beta}(\alpha,\beta)
\]
the conditional posterior for $p$ can evaluated using equation (\ref{N given p}) up to a constant, as follows: 
\begin{align*}
        p(p \mid \mu,\sigma,\mathbf{X},\mathbf{N}) & \propto p(\mathbf{X}\mid \mathbf{N},\mu ,\sigma,p)p(\mathbf{N}\mid p) p(p)\\
        & \propto \prod_{i=1}^m p(1-p)^{n_i-1} p^{a-1}(1-p)^{b-1}\\
        & \propto p^m(1-p)^{\sum_{i=1}^m n_i -m}p^{a-1}(1-p)^{b-1}\\
        & \propto p^{m+a-1}(1-p)^{\sum_{i=1}^m n_i -m+b-1}
\end{align*} 
Hence we get
\[ p(p \mid \mu, \sigma,X,N) \sim Beta(m+a,\sum_{i=1}^m n_i -m+b).
\]
Next for the mean and variance parameters $(\mu,\sigma^2)$ , consider the following prior:
\[\mu \mid \sigma^2 \sim N(v_0,\frac{\sigma^2}{n_0})\]
\[\sigma^2 \sim Inv-Gamma(\alpha,\beta)\]
From equation (\ref{X given others}), the likelihood of $p(\mathbf{X} \mid \mu,\sigma^2,p,\mathbf{N}), $ is:
\begin{align*}
        p(\mathbf{X} \mid \mu, \sigma^2,p, \mathbf{N}) & \propto \prod^m_{i=1} e^{-\frac{(x_i-n_i\mu)^2}{2n_i\sigma^2}}\\
        & \propto e^{-\frac{1}{2\sigma^2}\sum_{i=1}^m\frac{1}{n_i}(x_i-n_i\mu)^2}
\end{align*}
Again from equation (\ref{X given others}), we obtain conditional joint posterior of $(\mu,\sigma^2)$ as:
\begin{align*}
p(\mu,\sigma^2 \mid p,\mathbf{X},\mathbf{N}) & \propto p(X \mid \mu,\sigma,p,\mathbf{X},\mathbf{N}) p(\mu,\sigma^2)\\
& \propto \frac{1}{\sigma^m}e^{-\frac{1}{2\sigma^2}\sum^m_{i=1}\frac{1}{n_i}(x_i-n_i\mu)^2}\left(\frac{1}{\sigma^2}\right)^{\alpha +1 +1/2}e^{-\frac{1}{2\sigma^2}\left( 2\beta +  n_0(u-v_0)^2 \right)}\\
& \propto \left(\frac{1}{\sigma^2}\right)^{\frac{m}{2}+\alpha+1+\frac{1}{2}}e^{-\frac{1}{2\sigma^2}\left( 2\beta +  n_0(u-v_0)^2 + \sum^m_{i=1}\frac{1}{n_i}(x_i-n_i\mu)^2 \right)}
\end{align*}
For the conditional posterior of $\mu$, first note that, the prior on $\mu$ is,
\[p(\mu \mid \sigma^2) \propto e^{-\frac{n_0(\mu-v_0)^2}{2\sigma^2}}. \]  
Consequently we have,
\begin{align*}
    p(\mu \mid \sigma^2,p,\mathbf{X},\mathbf{N}) & \propto p(\mathbf{X} \mid \mu, \sigma^2,p,\mathbf{N})p(\mu \mid \sigma^2)\\
    & \propto e^{-\frac{1}{2\sigma^2}\sum_{i=1}^m\frac{1}{n_i}(x_i-n_i\mu)^2}.e^{-\frac{n_0(\mu-v_0)^2}{2\sigma^2}}\\
    & \propto e^{-\frac{1}{2\sigma^2}\left(\sum_{i=1}^m\frac{1}{n_i}(x_i^2+n_i^2\mu^2-2x_in_i\mu)+n_0(\mu-v_0)^2\right)}\\
    & \propto e^{-\frac{1}{2\sigma^2}\left(\sum_{i=1}^m\frac{x_i^2}{n_i}+ \mu^2\sum_{i=1}^m n_i -2\mu\sum_{i=1}^m x_i + n_0\mu^2 +n_0v_0^2 -2n_0v_0\mu  \right)}\\
    & \propto e^{-\frac{1}{2\sigma^2}\left( (\sum_{i=1}^m n_i +n_0)\mu^2 -2\mu(\sum_{i=1}^m x_i + n_0v_0) + \sum_{i=1}^m \frac{x_i^2}{n_i} +n_0v_0^2\right)}\\
    & \propto e^{-\frac{(\sum_{i=1}^m n_i +n_0)}{2\sigma^2}\left( \mu - \frac{n_0v_0 + \sum_{i=1}^m x_i}{\sum_{i=1}^m n_i +n_0}\right)^2}
\end{align*}
In the the last step, we have ignored terms not dependent on $\mu$ and completed the square. Hence
\[ \mu \sim N( \mu^*,\frac{\sigma^2}{n^*}), 
\]
where
\[ \mu^*=\frac{n_0v_0 + \sum_{i=1}^m x_i}{\sum_{i=1}^m n_i +n_0} \]
\[n^*=\sum_{i=1}^m n_i +n_0
\]

\subsubsection*{Conditional posterior distribution of $\sigma^2$ }
Since we have,
\[p(\mu \mid \sigma^2,p,X,N) \propto \left(\frac{1}{\sigma^2}\right)^{\frac{1}{2}}e^{-\frac{n^*(\mu-\mu^*)^2}{2\sigma^2}}
\]
The conditional posterior distribution of $\sigma$ obtained by dividing the joint posterior of $\sigma$ and $\mu$, by the full conditional posterior of $\mu$, does not depend on $\mu$, so we get,
\begin{align*}
    p(\sigma^2 \mid \mu,p,X,N) & \propto \frac{p(\mu,\sigma^2 \mid p,X,N)}{p(\mu \mid \sigma^2,p,X,N)}\\
    & \propto \left(\frac{1}{\sigma^2}\right)^{\frac{n}{2} + \alpha +1 +\frac{1}{2}-\frac{1}{2}}e^{-\frac{1}{2\sigma^2}\left( 2\beta +  n_0(u-v_0)^2 + \sum^m_{i=1}\frac{1}{n_i}(x_i-n_i\mu)^2 + n^{*}(\mu - \mu^{*})^2 \right)}
\end{align*}
The conditional posterior of $\sigma$ is independent of $\mu$, we compute the posterior by fixing $\mu=\mu^*$ we obtain,
\[  p(\sigma^2 \mid \mu,p,X,N)  \propto \left(\frac{1}{\sigma^2}\right)^{\frac{n}{2}+\alpha+1}e^{{-\frac{1}{2\sigma^2}\left( 2\beta +  n_0(u^*-v_0)^2 + \sum^m_{i=1}\frac{1}{n_i}(x_i-n_i\mu^*)^2 \right)}}
\]
Finally we obtain
\[ \sigma^2 \sim Inv-Gamma(\alpha^*,\beta^*)
\]
\[ \alpha^*= \alpha+ \frac{n}{2}  \]
\[ \beta^*= 2\beta +  n_0(u^*-v_0)^2 + \sum^m_{i=1}\frac{1}{n_i}(x_i-n_i\mu^*)^2 
\]
\subsubsection*{Conditional Posterior distribution of $N_i$}
In this case the conditional distribution is not in closed form , however we can still obtain the following proportionality,
\begin{align*}
        p(p \mid \mu,\sigma,\mathbf{X},\mathbf{N}) & \propto p(\mathbf{X}\mid \mathbf{N},\mu ,\sigma,p)p(\mathbf{N}\mid p)\\
        & \propto \prod^m_{i=1} \frac{1}{\sqrt{n_i}} e^{-\frac{(x_i-n_i\mu)^2}{2n_i\sigma^2}} (1-p)^{n_i-1}\\
\end{align*}
This gives us the distribution of each independent $N_i$
\subsection{Derivation of Varaitonal Update Equations}
Using the same notation, let $\mathbf{X}$ and $\mathbf{N}$ be the GSN random variables and  latent variables, respectively. Now, let the variational distribution be $q(N,p,\mu,\sigma^2)$ and assume the distribution of the latent variables and the parameters factorize as follows:
\[ q(N,p.\mu,\sigma^2) = q(N)q(p,\mu,\sigma^2)
\]
\subsubsection*{Variational Distribution of Latent Variables}
\begin{align}
\log q^*(N) &= E_{p,\mu,\sigma^2}(\log  p(X \mid N,\mu,\sigma^2,p)) + E_{p,\mu,\sigma^2}(\log  \pi(N \mid p)) + \mbox{const} \notag \\
&= E_{\mu,\sigma^2}\left(-\frac{1}{2\sigma^2}\sum_{i=1}^m\frac{1}{n_i}(x_i-n_i\mu)^2 - \sum_{i=1}^m \log \sqrt{n_i} - \frac{n}{2} \log \sigma^2\right) + \notag \\
& \qquad E_p\left(\left(\sum_{i=1}^m n_i - n\right)\log p(1-p)\right) + \mbox{const} \notag \\
&= -\sum_{i=1}^m \log \sqrt{n_i}  + E_p(\log (1-p)) \left(\sum_{i=1}^m n_i\right)  - \notag \\ 
& \qquad E_{\mu, \sigma^2}\left(\frac{1}{2\sigma^2}\sum_{i=1}^m\frac{1}{n_i}(x_i-n_i\mu)^2\right)  + \mbox{const}
\label{VI for latent}
\end{align}
The expectation with respect to the parameters can be found by using the Variational posterior distribution of the parameters derived in the next section
\subsubsection*{Variational Distribution of Parameters}
The Variational distribution of $p,\mu,\sigma^2$ is given as,
\begin{align}
\log q^*(p,\mu,\sigma^2)&= E_N(\log p(\mathbf{X} \mid \mathbf{N},\mu,\sigma^2,p)) + E_N(\log  p(N \mid p)) + E_N(\log \pi(p)) + \notag \\
& \qquad E_N(\log p(\mu,\sigma^2))
\label{VI posterior}
\end{align}
We can further factorise $q(p,\mu,\sigma^2)$ as $q(p)q(\mu, \sigma^2)$ and  $q(\mu, \sigma^2)$ as $q(\mu \mid \sigma^2)q(\sigma^2)$.
Using the prior assumption on p and (\ref{VI posterior}) we have
\begin{align*}
    \log q^*(p)&=E_N(\log \pi(N \mid p))+ E_N(\log \pi(p)) + \mbox{const}\\
    &= n\log p + \left(E_N\left(\sum_{i=1}^m n_i\right)-n\right) \log (1-p) + (a-1)\log (p) + \\
    & \qquad (b-1)\log (1-p) +\mbox{const}\\
    &= (n+a-1)\log p + (E_N(\sum_{i=1}^m n_i)-n+ b-1)\log (1-p) +\mbox{const},
\end{align*}
which in turn implies that
\[ q^*(p) = \mbox{Beta}\left(n+a,E_N\left(\sum_{i=1}^m n_i\right)-n+b\right)
\]
Further, for joint density of $\mu,\sigma^2$, we have 
\[ \log q^*(\mu,\sigma^2)=\log q^*(\mu \mid \sigma^2) + \log q^*(\sigma^2)
\]
Similar to the derivation of the conditional posterior distributions we once again use the conjugacy of our prior assumptions and (\ref{VI posterior}) to obtain,
\[ q^*(\mu \mid \sigma^2) = N(\mu^*,\frac{\sigma^2}{n^*}),
\]
where
\[ \mu^*=\frac{\sum_{i=1}^m x_i + n_0v_0}{n_0+E_N\bigg( \sum_{i=1}^m n_i \bigg)}
\]
\[n^*=n_0 + E_N(\sum_{i=1}^m n_i).
\]
For $\sigma^2$ we use the following relation
\begin{align*}
    \log q^*(\sigma^2 ) &= \log q^*(\mu,\sigma^2) - \log q^*(\mu \mid \sigma^2)\\
\end{align*}
Upon simplifying the above expression, we get 
\[ q^*(\sigma^2)=Inv-Gamma(\alpha^*,\beta^* ),
\]
where
\[ \alpha^*=\alpha +1+\frac{n}{2}, \text { and }
\]
\[ \beta^*=2 \beta  + n_0(\mu^*-v_0)^2 + E_N\left( \sum_{i=1}^m \frac{1}{n_i}(x_i-n_i\mu^*)^2\right).
\]
\subsubsection*{Evidence Lower Bound (ELBO) for GSN distribution}
\[ ELBO(q)= \sum_{N} \int \int\int q(\textbf{N},p,\mu,\sigma^2)log \{ \frac{p(\textbf{X},\textbf{N},p,\mu,\sigma^2)}{q(\textbf{N},p, \mu,\sigma^2)} \} dp d\mu d\sigma^2 
\]
We obtain the final expression as,
 \begin{align*}
     ELBO(q) &= E \left[ log \, p(\textbf{X},\textbf{N},p,\mu,\sigma^2)\right] - E\left[log \, q(\textbf{N},p, \mu,\sigma^2) \right]\\
     & = E \left[ p(\textbf{X} \mid \textbf{N},\mu,\sigma^2)\ \right] + E \left[ log \, p(\textbf{N}\mid p)\right] + E \left[log \, p(p) \right] + E \left[log \, p(\mu,\sigma^2) \right] \\ 
      & \quad - E \left[ log \, q(\textbf{N}) \right]
     - E \left[ log \, q(p)\right]- E \left[ log \, q(\mu, \sigma^2) \right]
 \end{align*}
All expectations are taken with respect to the variational distribution of the respective parameters $(q)$. 
\subsection{Informative Bayesian Inference for ASN}
We briefly describe the Gibbs sampling algorithm that we have used for the comparison study. For the full details of Bayesian Inference for ASN with informative priors we refer to \cite{informative}. 
Using the same notation as \cite{informative}, the pdf of ASN is given as
\[ f(x)=\frac{2}{\omega}\phi(\frac{x-\xi}{\omega})\Phi(\alpha\frac{x-\xi}{\omega} )\qquad x \in \bm{R}
\]
We have denoted the mean by $\xi$, variance by $\omega^2$ and skewness parameter by $\alpha$ (Earlier denoted as $\mu, \sigma^2$ and  $\lambda$)

Following the methods of \cite{Bayes}, and \cite{last}, relying on the stochastic representation of the ASN
distribution introduced by \cite{ASN},  we introduce independent standard
normal latent variables $\eta_1,\ldots, \eta_n$. Conditionally on such latent variables, we
can consider the i-th observation as being normally distributed with
mean  $\delta(y_i-\xi)$ and variance $ w^2(1-\delta^2) $. This interpretation lends to a conjugacy for conditional posterior of the parameters $(\xi, \omega^2)$ for the following choice of priors:
\[ (\xi,\omega^2) \sim N-\Gamma(\xi_0, \kappa \omega^2, a, b)
\]
We assume an ASN prior for the skewness parameter, $\alpha$
 \[
    \alpha \sim ASN(\alpha_0,\psi_0,\lambda_0)
 \]
\cite{informative} have shown that the conditional posterior , $\pi(\alpha \mid y)$ , belongs to the Unified Skew Normal class of distributions or SUN. (\cite{unified}). 

Consider the iid ASN sample $y=(y_1, \ldots, y_n)$ and we obtain the conditional posterior $\pi(\alpha \mid y)$
\begin{align}
    \label{skewness posterior}
    \alpha \mid y \sim SUN_{1,n+1}(\alpha_0, \gamma_2, \psi_0, 1, \Delta_2, \Gamma_2),
\end{align}
where $\Delta_2= [\delta_i]_{i=1,...,n+1}$ with $\delta_i=\psi_0z_i(\psi_0^2z_i^2+1)^{-1/2}$ and $z=(y^T,\lambda_0\psi_0^{-1})^T$, $\gamma_2=(\Delta_{2;1:n}\alpha_0\psi_0^{-1},0)$, $\Gamma_2=\mathcal{I} - D(\Delta_2) + \Delta_2\Delta_2^T.$

For prior elicitation we once again refer to \cite{informative}
\subsubsection*{Gibbs Sampling Algorithm}
\begin{enumerate}
    \item Update $\eta_i$ from it full conditional posterior distribution \[
    \eta_i \sim TN_0(\delta(y_i-\xi),w^2(1-\delta^2))
    \]
    where $\delta $ is $\alpha/\sqrt{\alpha^2+1}$ and $TN_{\tau}(\mu,\sigma^2)$ is mean $\mu$, variance $\sigma^2$ normal truncated below $\tau$
   \item Sample \[
          \xi   \sim N(\hat{\mu}, \hat{\kappa}\omega^2)  \]\[
          \omega^2 \mid \xi \sim Inv-Gmma(a+(n+1)/2, b + \hat{b}) 
   \]
    where
    \begin{align*}
    \widehat{\mu} & = \frac{\kappa\sum_{i=1}^n (y_i - \delta\eta_i) + (1-\delta^2)\xi_0}{n\kappa + (1-\delta^2)}, \hat{\kappa} = \frac{\kappa(1-\delta^2)}{n\kappa + (1-\delta^2)}\\
        \hat{b}& =\frac{1}{2(1-\delta^2)} \left( \delta^2\sum_{i=1}^n \eta_i^2\- 2\delta\sum_{i=1}^n \eta_i (y_i - \xi) + \delta\sum_{i=1}^n (y_i - \xi) + \frac{1-\delta^2}{\kappa} (\xi -\xi_0)^2 \right).
    \end{align*}
    \item Sample $\alpha$ from 
    \[ \alpha \sim \pi(\alpha \mid y^*)
    \]
    where  $y_i^*= (y_i - \xi)/\omega$  for  $i=1\ldots n$ and $\pi(\alpha \mid y)$ is given in equation (\ref{skewness posterior}).
\end{enumerate}
\bibliographystyle{apalike}
\bibliography{protoRefs}

\begin{thebibliography}{}

\bibitem[AAzzalini and Valle, 1996]{masn}
AAzzalini, A. and Valle, A.~D. (1996).
\newblock {The multivariate skew-normal distribution}.
\newblock {\em Biometrika}, 83(4):715--726.

\bibitem[Adcock et~al., 2015]{actuarial}
Adcock, C., Eling, M., and Loperfido, N. (2015).
\newblock Skewed distributions in finance and actuarial science: a review.
\newblock {\em The European Journal of Finance}, 21(13-14):1253--1281.

\bibitem[Arellano-Valle and Azzalini, 2006]{unified}
Arellano-Valle, R.~B. and Azzalini, A. (2006).
\newblock On the unification of families of skew-normal distributions.
\newblock {\em Scandinavian Journal of Statistics}, 33(3):561--574.

\bibitem[Arellano-Valle et~al., 2009]{last}
Arellano-Valle, R.~B., Genton, M.~G., and Loschi, R.~H. (2009).
\newblock Shape mixtures of multivariate skew-normal distributions.
\newblock {\em Journal of Multivariate Analysis}, 100(1):91--101.

\bibitem[Arnold and Beaver, 2000]{skewcauchy}
Arnold, B.~C. and Beaver, R.~J. (2000).
\newblock The skew-cauchy distribution.
\newblock {\em Statistics \& Probability Letters}, 49(3):285--290.

\bibitem[Azzalini, 1985]{ASN}
Azzalini, A. (1985).
\newblock A class of distributions which includes the normal ones.
\newblock {\em Scandinavian Journal of Statistics}, 12(2):171--178.

\bibitem[Azzalini et~al., 2014]{comparison1}
Azzalini, A., Browne, R.~P., Genton, M.~G., and McNicholas, P.~D. (2014).
\newblock Comparing two formulations of skew distributions with special
  reference to model-based clustering.
\newblock {\em arXiv preprint arXiv:1402.5431}.

\bibitem[Bayes and Branco, 2007]{Bayes}
Bayes, C. and Branco, E. (2007).
\newblock Bayesian inference for the skewness parameter of the scalar
  skew-normal distribution.
\newblock {\em Brazilian Journal of Probability and Statistics}, 21:141--163.

\bibitem[Bishop, 2006]{Bishop}
Bishop, C.~M. (2006).
\newblock {\em Pattern Recognition and Machine Learning (Information Science
  and Statistics)}.
\newblock Springer-Verlag, Berlin, Heidelberg.

\bibitem[Blei et~al., 2017]{VI}
Blei, D.~M., Kucukelbir, A., and McAuliffe, J.~D. (2017).
\newblock Variational inference: A review for statisticians.
\newblock {\em Journal of the American Statistical Association},
  112(518):859--877.

\bibitem[Canale and Scarpa, 2013]{informative}
Canale, A. and Scarpa, B. (2013).
\newblock Informative bayesian inference for the skew-normal distribution.

\bibitem[Eling, 2012]{comparison2}
Eling, M. (2012).
\newblock Fitting insurance claims to skewed distributions: Are the skew-normal
  and skew-student good models?
\newblock {\em Insurance: Mathematics and Economics}, 51(2):239--248.

\bibitem[Fasano et~al., 2019]{vskew}
Fasano, A., Durante, D., and Zanella, G. (2019).
\newblock Scalable and accurate variational bayes for high-dimensional binary
  regression models.

\bibitem[Figueiredo and Gomes, 2013]{SPC}
Figueiredo, F. and Gomes, M. (2013).
\newblock The skew-normal distribution in spc.
\newblock {\em Revstat - Statistical Journal}, 11:83--104.

\bibitem[Ghaderinezhad et~al., 2020]{Review}
Ghaderinezhad, F., Ley, C., and Loperfido, N. (2020).
\newblock Bayesian inference for skew-symmetric distributions.
\newblock {\em Symmetry}, 12:491.

\bibitem[Hossain and Beyene, 2015]{bio}
Hossain, A. and Beyene, J. (2015).
\newblock Application of skew-normal distribution for detecting differential
  expression to microrna data.
\newblock {\em Journal of Applied Statistics}, 42(3):477--491.

\bibitem[Jones and Faddy, 2003]{skewt}
Jones, M.~C. and Faddy, M.~J. (2003).
\newblock A skew extension of the t-distribution, with applications.
\newblock {\em Journal of the Royal Statistical Society. Series B (Statistical
  Methodology)}, 65(1):159--174.

\bibitem[Kazemi and Noorizadeh, 2015]{comparison3}
Kazemi, R. and Noorizadeh, M. (2015).
\newblock A comparison between skew-logistic and skew-normal distributions.
\newblock {\em MATEMATIKA: Malaysian Journal of Industrial and Applied
  Mathematics}, 31(1):15–24.

\bibitem[Kundu, 2014]{GSN}
Kundu, D. (2014).
\newblock Geometric skew normal distribution.
\newblock {\em Sankhyā: The Indian Journal of Statistics, Series B (2008-)},
  76(2):167--189.

\bibitem[Kundu, 2017]{mgsn}
Kundu, D. (2017).
\newblock Multivariate geometric skew-normal distribution.
\newblock {\em Statistics}, 51(6):1377--1397.

\bibitem[Ley, 2013]{skew}
Ley, C. (2013).
\newblock {\em Skew Distributions}.
\newblock American Cancer Society.

\bibitem[Liseo and Loperfido, 2006]{reference}
Liseo, B. and Loperfido, N. (2006).
\newblock A note on reference priors for the scalar skew-normal distribution.
\newblock {\em Journal of Statistical Planning and Inference}, 136:373--389.

\bibitem[Nadarajah, 2009]{skewlogistic}
Nadarajah, S. (2009).
\newblock The skew logistic distribution.
\newblock {\em AStA Advances in Statistical Analysis}, 93(2):187--203.

\bibitem[Redivo et~al., 2020]{clustering}
Redivo, E., Nguyen, H.~D., and Gupta, M. (2020).
\newblock Bayesian clustering of skewed and multimodal data using geometric
  skewed normal distributions.
\newblock {\em Computational Statistics \& Data Analysis}, 152:107040.

\bibitem[Simon, 1955]{skeweg}
Simon, H.~A. (1955).
\newblock On a class of skew distribution functions.
\newblock {\em Biometrika}, 42(3/4):425--440.

\bibitem[Wang et~al., 2004]{gskew}
Wang, J., Boyer, J., and Genton, M.~G. (2004).
\newblock A skew-symmetric representation of multivariate distributions.
\newblock {\em Statistica Sinica}, 14(4):1259--1270.

\end{thebibliography}
\end{document}